\documentclass[useAMS,usenatbib]{mn2e}
\usepackage{latexsym}
\usepackage{makeidx}
\usepackage{cite}
\usepackage{amsmath}
\usepackage{color}
\usepackage{upgreek}
\usepackage{morefloats}
\usepackage{lineno}
\usepackage{setspace}
\usepackage{graphicx}

\title[Transit spectroscopy with JWST: Systematics]{Transit spectroscopy with JWST: Systematics, starspots and stitching}
\author[J.~K. Barstow et al.]{J.~K. Barstow$^{1,2}$\thanks{E-mail:
jo.barstow@astro.ox.ac.uk (JKB)}, S. Aigrain$^{1}$,
P.~G.~J. Irwin$^{2}$, S. Kendrew$^{1}$, L.~N. Fletcher$^{2}$\\
$^{1}$Astrophysics, Denys Wilkinson Building, University of Oxford, UK\\
$^{2}$Atmospheric, Oceanic and Planetary Physics, Clarendon Laboratory, University of Oxford, UK}
\begin{document}

\date{Submitted *** 2014}

\pagerange{\pageref{firstpage}--\pageref{lastpage}} \pubyear{2014}

\maketitle

\label{firstpage}

\begin{abstract}
The \textit{James Webb Space Telescope (JWST)} is predicted to make great advances in the field of exoplanet atmospheres. Its 25 m$^2$ mirror means that it can reach unprecedented levels of precision in observations of transit spectra, and can thus characterise the atmospheres of planets orbiting stars several hundred pc away. Its coverage of the infrared spectral region between 0.6 and 28 $\upmu$m allows the abundances of key molecules to be probed during the transit of a planet in front of the host star, and when the same planet is eclipsed constraints can be placed on its temperature structure. In this work, we explore the possibility of using low-spectral-resolution observations by \textit{JWST}/NIRSpec and \textit{JWST}/MIRI-LRS together to optimise wavelength coverage and break degeneracies in the atmospheric retrieval problem for a range of exoplanets from hot Jupiters to super Earths. This approach involves stitching together non-simultaneous observations in different wavelength regions, rendering it necessary to consider the effect of time-varying instrumental and astrophysical systematics. We present the results of a series of retrieval feasibility tests examining the effects of instrument systematics and star spots on the recoverability of the true atmospheric state, and demonstrate that correcting for these systematics is key for successful exoplanet science with \textit{JWST}. 
\end{abstract}


\maketitle 

\section{Introduction}
The field of exoplanet atmospheric science has expanded rapidly since the first transit spectroscopy measurement of the hot Jupiter HD 209458b's atmosphere by \citet{charb02}. Since then, the atmospheres of many hot Jupiters and some smaller planets have been at least partially characterised using this technique; the presence of clouds is suggested by observations of HD 189733b (e.g. \citealt{pont13}), WASP-12b (e.g. \citealt{sing13,stevenson14}) and GJ 1214b (e.g. \citealt{kreidberg14}); H$_2$O absorption features have been measured for several hot Jupiters using the \textit{Hubble Space Telescope}/WFC (e.g. \citealt{deming13,wakeford13,mandell13,fraine14}); thermal emission spectra and phase curves have been obtained for HD 189733b and HD 209458b (e.g. \citealt{deming06,charb08,knutson12,zellem14}); and the visible reflection spectrum of HD 189733b has also been measured \citep{evans13}. 

\begin{figure}
\centering
\includegraphics[width=0.5\textwidth]{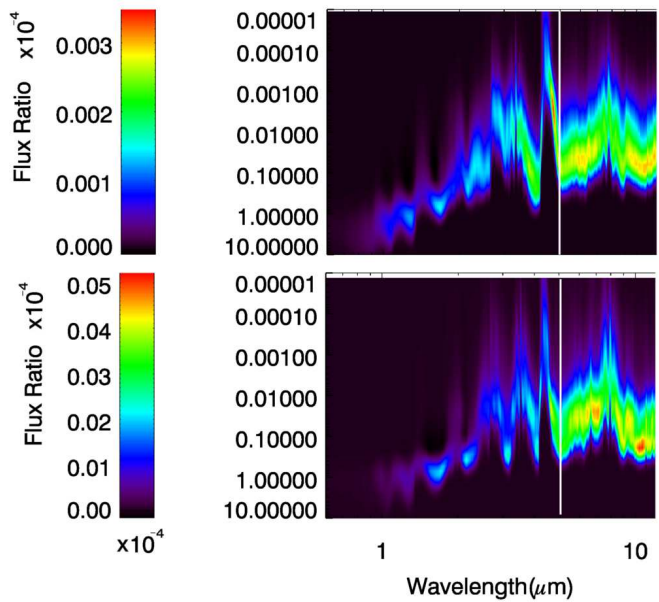}
\caption{Retrieval sensitivity to temperature as a function of pressure and wavelength, for a hot Jupiter orbiting a sun-like star (top) and a hot Neptune orbiting an M dwarf (bottom). The contours correspond to a change in observed flux ratio for a 1 K change in the temperature at each level in the atmosphere. The division between the NIRSpec and MIRI instruments is marked as a thin white line. MIRI has sensitivity over a very limited pressure range, but has higher sensitivity than NIRSpec over that range.\label{tempcov}}
\end{figure}

Transit spectroscopy is a technique used to infer the atmospheric properties of exoplanets, when the planet transits or is eclipsed by its host star. During a planetary transit, opacity sources in the atmosphere can make the planet appear larger at particular wavelengths, allowing atmospheric composition to be inferred. During eclipse, or secondary transit, the planet's thermal emission spectrum can be extracted by comparing the in- and out-of-transit fluxes, providing information about atmospheric structure. The fluctuations in the transit depth as a function of wavelength are small, of order 1/10,000 for a hot Jupiter-size planet and as small as 1/1000,000 for Earthlike worlds. The infrared\textit{James Webb Space Telescope (JWST)} will transform our ability to study exoplanet atmospheres, using transit spectroscopy and other methods; for example, its large collecting area and coronographic capability will make it an extremely useful tool for direct spectroscopic imaging of hot, young exoplanets far from their host stars, such as the HR 8799 system. In addition, it will greatly expand the distance out to which spectroscopic follow up is possible for transiting exoplanets, due to its large collecting area, although its high sensitivity limits the possibility of investigating closer planets as their host stars may be bright enough to saturate one of the instruments. Given the size of the transit depth fluctuations that are the measurement of interest, \textit{JWST}'s large collecting area and high sensitivity are beneficial to increase the signal to noise ratio. Instrument systematics are also likely to play a significant role, and the low photon noise should make these easier to identify and therefore allow us to reach the required high precision. \textit{JWST} is the first space telescope to be designed and tested in the era of exoplanet spectroscopy, and slight modifications were made to optimise the telescope for that purpose; we also have the opportunity to study and find ways of correcting for systematics prior to launch, and \textit{JWST}'s intended location at the L2 point should minimise thermal fluctuations.

The main difficulties in transit spectroscopy have arisen because available instruments generally only cover a narrow wavelength range, meaning that spectra taken at different times from the ground or from space must be stitched together to obtain a coherent picture of the planet's atmosphere. Broad wavelength coverage is particularly useful for breaking degeneracies between gas abundances and temperature structure in secondary transit (\citealt{barstow13}, Figure~\ref{tempcov}) and gas abundances and cloud coverage in primary transit \citep{barstow13b}, because spectral properties of clouds vary slowly as a function of wavelength.  Problems arise when stitching together non-simultaneous measurements of transits because the recorded baseline flux of the star may change between observations, due to, for example, stellar activity or variation in systematics between instruments/within a single instrument over time. This can lead to obvious mismatches between overlapping spectra (e.g. FORS observations of GJ 1214b by \citealt{bean11}) or the spurious detection of atmospheric features (e.g. the GJ 1214b Ks-band observation by \citealt{croll11}, initially interpreted as evidence of methane absorption in a low mean molecular weight atmosphere but not confirmed by the narrow band observations of \citealt{colon13}). \textit{JWST} will eliminate a great deal of the need for stitching due to the relatively broad wavelength coverage of some of its instruments. The broadest near-simultaneous wavelength coverage with JWST can be achieved by combining a near-infrared instrument with the Mid-Infrared Instrument (MIRI, \citealt{reike15}). MIRI offers two spectroscopic observing modes: the low-resolution prism mode (LRS), which covers 5---12 $\upmu$mm at R$\sim$100 \citep{kendrew14}; and the medium-resolution integral field spectrometer (MRS), which extends from 5 to 28 $\upmu$mm at R$\sim$1500-3500. While the MRS provides greater wavelength coverage, the LRS is more suitable for exoplanet transit spectroscopy, as it covers the 5---12 $\upmu$m region in a single exposure. The MRS requires three separate exposures to cover the full range. The LRS can in addition be used in slitless mode which is optimised for observations of bright compact sources such as exoplanet host stars, and our study assumes use of this mode.

Of the near-infrared instruments, the Near-Infrared Camera (NIRCam) and the Near-Infrared Imager and Slitless Spectrograph (NIRISS) will both be valuable for exoplanet observations. However, given the importance of wavelength coverage for our work, we focus instead on the Near-Infrared Spectrograph (NIRSpec), which provides the largest single-shot wavelength coverage of all near-infrared instruments with its fixed-slit low-resolution mode. Like MIRI, NIRSpec offers a higher-resolution integral field spectroscopy mode, but this cannot provide the full 0.6---5.0 $\upmu$mm in a single exposure. The low-resolution prism mode uses fixed slits.

Since the Near InfraRed Spectrometer (NIRSpec) and MIRI cannot be used simultaneously, accurate stitching of spectra will be important for the success of \textit{JWST} transit observations. NIRSpec in prism mode, covering 0.6---5 $\upmu$m, and MIRI's Low Resolution Spectrograph in slitless mode, covering 5---12 $\upmu$m, provide the largest wavelength range possible with only two separate observations and in this work we focus on these instrument modes. However, our study of instrumental offsets and starspots will have implications for combining measurements from other instruments too.

In Figure~\ref{tempcov} we show the sensitivity to temperature at different pressure levels in the atmosphere with NIRSpec and MIRI, for a hot Jupiter orbiting a sun-like star (top) and a hot Neptune orbiting an M dwarf (bottom). At each wavelength, the maximum sensitivity occurs in the region of the atmosphere the radiation emerges from. Sensitivity over a wide pressure range is required to perform spectral inversions and obtain the atmospheric structure, and this is best facilitated by broad wavelength coverage. For example, the MIRI instrument has little sensitivity in the deep atmosphere, meaning that the temperature structure could not be fully retrieved using MIRI alone. Whilst NIRSpec does have sensitivity over a wide range of pressures, including MIRI provides further gas absorption bands, which can help to resolve any remaining degeneracies, and in secondary transit the flux ratio increases as a function of wavelength so the planetary signal is stronger in the MIRI range. We therefore consider in this paper a combination of NIRSpec and MIRI observations. 

\begin{table}
\centering
\begin{tabular}[c]{|c|c|c|}
\hline
Planet & Mass ($\times$10$^{24}$ kg) & Radius (km)\\
\hline
Hot Jupiter & 1800 & 75000 \\
Hot Neptune & 180 & 30000 \\
GJ 1214b & 37 & 15355\\
`Earth' & 6 & 6378 \\
\hline
\end{tabular}
\caption{Masses and radii of the model planets in this study.\label{planet_properties}}
\end{table}

As mentioned above, NIRISS and NIRCam could also be used for transit spectroscopy. They both cover a similar spectral range to that of NIRSpec, but neither can simultaneously observe the 0.6---5 $\upmu$m range, which would introduce further complexity when stitching together observations to make a complete spectrum. However, these instruments do have potential advantages over NIRSpec as their saturation limits are less strigent, so they are likely to be useful for observing closer, brighter systems. The bright target limit for NIRSpec in prism mode is a host star J-band magnitude of ~9.75 \citep{ferruit14}, equivalent to GJ 1214, but brighter targets such as HD 189733 can be reached using the higher resolving power modes. In addition, the higher resolving power mode for MIRI is likely to be useful. The MIRI integral field unit extends out to 28 $\upmu$m at R$\sim$3000, which will allow high resolution observations of, for example, the 15 $\upmu$m CO$_2$ band, but use of this mode introduces further difficulty with stitching as the observable wavelength range is split into three segments \citep{wells15}. 

In this work, we use the NEMESIS radiative transfer and retrieval code \citep{irwin08} and follow the method adopted by \citet{barstow13} to test the accuracy with which combined \textit{JWST}/NIRSpec and MIRI observations can recover the true atmospheric state of a range of planets, described in Section~\ref{modelplanets}. To explore the effect of instrumental and astrophysical systematic noise, we simulate the effect of baseline flux offsets between NIRSpec and MIRI, and also the effect of time varying stellar activity on temporally disconnected observations (Section~\ref{noise}). The method and the NEMESIS code are briefly described in Section~\ref{code} and our results are presented in Section~\ref{results}, with discussion of future work and suggestions for appropriate strategies in Section~\ref{discussion}.

\section{Model Planets and Synthetic Spectra}
\label{modelplanets}

\begin{table*} \centering \begin{tabular}[c]{|c|c|} \hline Gas & Source\\ \hline
H$_2$O & HITEMP2010 \citep{roth10}\\ CO$_2$ &  CDSD-1000 \citep{tash03}\\ CO &
HITRAN1995 \citep{roth95}\\ CH$_4$ & STDS \citep{weng98}\\H$_2$/He & \citep{borysow89,borysowfm89,borysow90,borysow97,borysow02}\\ \hline \end{tabular} \caption{Sources
of gas absorption line data.\label{gasdata}} \end{table*} 

The model planets we consider here are similar to those studied in \citet{barstow13} for the Exoplanet Characterisation Observatory (EChO) ESA M3 mission proposal. The bulk properties are listed in Table~\ref{planet_properties}. So far, we focus on Jupiter and (mini) Neptune-sized gas giant planets, which are considerably simpler to model as they are assumed to be dominated by H$_2$ and He. Observation of super Earths with \textit{JWST}/NIRSpec has been explored by \citet{benneke12} and \citet{dewit13}, and the major difficulty is determining the bulk atmospheric composition from transmission spectra since this is degenerate with temperature and cloud properties. We defer further exploration of this to future work, although we briefly discuss the possibility of detecting O$_3$ on an Earth analogue planet in orbit around a nearby M dwarf.  

As in \citet{barstow13}, we use NEMESIS to generate a series of synthetic spectra based on this range of model planets; a hot Jupiter-size planet orbiting a Sun-like star (radius 695000 km), a hot Neptune/Earth analogue orbiting an M dwarf (radius 97995 km, equal to that of Proxima Centauri) and a set of spectra based on the super-Earth GJ 1214b.  The hot Jupiter and hot Neptune don't bear any direct relationship to observed planets, but the hot Jupiter may be thought of as an HD 209458b/HD 189733b analog, albeit in orbit around a slightly hotter star. These planets both have clear atmospheres, but we include cloud in the GJ 1214b and Earth analog cases: the GJ 1214b cloud is assumed to be made of Titan tholins and has two particle populations with effective radii of 0.1 and 1 $\upmu$m (for further details see \citealt{barstow13b}); the Earth analog cloud is a water cloud parameterised as in \citet{irwin14}. We defer a more detailed exploration of different cloudy atmospheres to a future publication.  We use the Kurucz solar model spectrum\footnotemark for the Sun-like star; this is located at 250 pc, significantly further away than the equivalent case in \citet{barstow13}, because \textit{JWST}'s high sensitivity and large mirror relative to EChO mean that NIRSpec in prism mode can be easily saturated \citep{ferruit14}. PHOENIX models of the appropriate temperature, log($g$) and metallicity are used for the Proxima Centauri-based M dwarf and GJ 1214 \citep{husser13}. The hot Neptune-M dwarf system is located 15 pc away, and the Earth-M dwarf system 10 pc away. GJ 1214 is at its correct distance of 14.6 pc. 

\begin{figure*}
\centering
\includegraphics[width=0.85\textwidth]{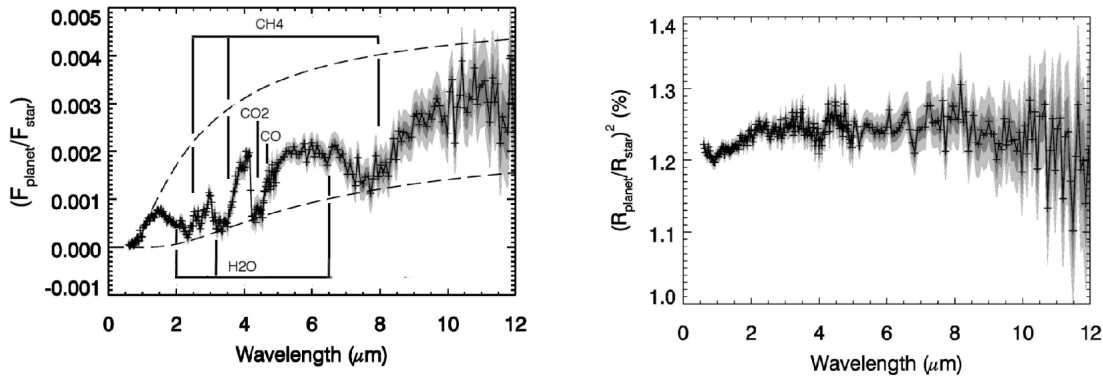}
\caption{Example spectra for the hot Jupiter orbiting a sun-like star. Left: a single secondary transit (dashed lines indicate the blackbody flux ratios at 2500 K, top, and 1200 K, bottom); right: a single primary transit. Dark/light grey shading indicate 1$\sigma$/2$\sigma$ error bars.\label{hotjup_spectra}}
\end{figure*}

For each planet, we keep the T-p profile and bulk atmospheric composition (N$_2$-O$_2$ for Earth, H$_2$-He for all other planets) constant, but we vary the trace gas abundances for each separate secondary transit retrieval. The atmospheres are assumed to be well-mixed in all cases except the Earth analog, which has gas abundance vertical profiles as measured on Earth \citep{irwin14}. For the primary transit retrieval we also vary the radius at the 10-bar pressure level, as primary transit retrievals are extremely sensitive to the precise size of the planet. In the GJ 1214b model runs the optical depths of the two clouds are also allowed to vary. Since our aim is to explore the capability of \textit{JWST} to characterise a GJ 1214b-like planet, the variations on the GJ 1214b model that we use are not required to be compatible with the existing observed spectrum, although they are all similar to the best-fit case from \citet{barstow13b}.  

 \subsection{T-p Profiles}
The T-p profiles used in this work (with the exception of the Earth analog case, which uses an average Earth profile) are based on a single-slab atmosphere model (Equation~\ref{slab}). We assume that the atmospheric temperature decreases adiabatically through the troposphere and is either isothermal above that level or includes a temperature inversion. The model troposphere extends up to 0.1 bar, the approximate location of the tropopause for the solar system giant planets, and stars at the 1 bar level. There is very little sensitivity to temperature below the 1 bar level (see Figure~\ref{tempcov}) and so we leave the temperature below this level constant.

\begin{equation}
T_{\mathrm{trop}}=T_{\mathrm{strat}} -\Gamma\frac{kT_{\mathrm{strat}}}{mg}\mathrm{ln}\left(\frac{p_1}{p_2}\right) \label{slab}
\end{equation}
$T_{\mathrm{trop}}$ and $T_{\mathrm{strat}}$ are the temperatures at the bottom and top of the adiabatic region, $\Gamma$ is the dry adiabatic lapse rate, $k$ is the Boltzmann constant, $m$ is the molar weight of the atmosphere, $g$ is the gravitational acceleration and $p_1$ and $p_2$ are the pressures at the bottom and top of the adiabatic region, in this case 1 and 0.1 bar.

Whilst these temperature profiles are influenced only by very basic physics, they are sufficient to test the retrievability of temperature and composition with \textit{JWST}. To increase the range of temperature profiles tested, we include some with a temperature inversion (a region of temperature increase with altitude) above the tropopause. For a real planet, this may be caused by the presence of an ultraviolet or visible absorber, such as ozone in the case of the Earth or metal oxides in the case of hot Jupiters. In the test case here, it is simply included to ensure that an inversion would be detectable should it be present.

Throughout this work, we refer to: the input temperature profile, which is the profile used to generate the noisy synthetic spectra; the \textit{a priori} temperature profile, which is the profile used by \textit{NEMESIS} as the starting point for the retrieval; and the retrieved temperature profile, which is the profile that \textit{NEMESIS} determined to provide the best fit to the noisy synthetic spectrum. If the retrieval is accurate, the retrieved temperature profile should closely match the input temperature profile and should not be influenced by the \textit{a priori} temperature profile.

\footnotetext{http://kurucz.harvard.edu/stars/}

\subsection{Noise Calculation}
\label{noise}
The noise calculation in this work is similar to that of \citet{barstow13}; we assume random noise at the photon
 limit for both instruments, using the appropriate values for total instrument throughput for NIRSpec and MIRI in the chosen observational modes. We refer the reader to \citet{barstow13} for details of the noise calculation. The instrument properties used are as follows: for NIRSpec, we adopt the average transmission and quantum efficiency properties used by \citet{deming09}, assuming a detector QE of 0.8, a telescope optics efficiency of 0.88 and a total NIRSpec optics transmission of 0.4; for MIRI-LRS, we use the photon conversion efficiencies presented in \citet{kendrew14}, which represents the fraction of photons from the source that are eventually received and recorded by the detector. 

In addition to the photon noise calculated for these instruments, we also include simulated offsets due to instrument systematics and due to star spots. For the instrument systematics, we calculated the spectrally-averaged photon noise for each instrument, and generate a series of Gaussian random numbers (mean 0, standard deviation 1). We then add a number from this series, multiplied by half the NIRSpec-averaged photon noise, to all NIRSpec points, and subtract the same number multiplied by half the MIRI-averaged photon noise from the MIRI points. This results in a range of models for which the NIRSpec and MIRI spectra are offset from each other by a random amount on the order of the photon noise; for half the cases, the NIRSpec spectrum will be shifted above the MIRI spectrum and otherwise it will be below the MIRI spectrum. We also repeated the test using 3$\times$ photon noise, and we performed this test for both primary and secondary transit observations. 

\subsection{Star Spots}
Star spots do not have a significant effect on secondary transit observations, but can alter primary transit spectra. This occurs because, if the star has spots on its disc, the brightness of the transit chord may not be the same as the disc-averaged brightness. Following \citet{pont13}, if the transit chord is more spotty than the disc average, then the planet radius is underestimated because the planet is crossing a less bright part of the star; conversely, if the transit chord crosses a spot-free region, the radius is overestimated because the occulted region is brighter than the average. The first case can be easier to identify, as the signatures of spot crossing events may be seen in the transit lightcurve, although the wavelength ranges of  MIRI and possibly NIRSpec are too red for these events to be seen above the noise since the spot contrast decreases towards longer wavelengths. In the second case, because no spots are crossed, there is not necessarily any evidence that the star has active regions when the transit is observed, so any potential errors introduced by stellar activity may be unidentified. Therefore, we focus on the second case here.
\begin{figure}
\centering
\includegraphics[width=0.5\textwidth]{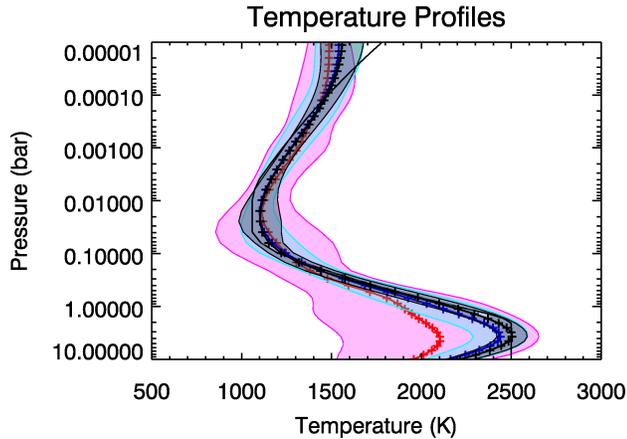}
\caption{Temperature retrievals and error envelopes for the hot Jupiter secondary transit; cases shown are no baseline offset between NIRSpec and MIRI (black/grey crosses, lines and shading), offset of order of the photon noise (blue/aqua) and offset of order 3$\times$ the photon noise (red/pink). The input temperature profile is shown by the solid black line. In all cases, the \textit{a priori} temperature profile is an isotherm at 1500 K. The retrieved temperatures shown are an average over 50 retrievals with different atmospheric compositions, and the envelope is the standard deviation of these.\label{hotjuptempcompare}}
\end{figure}

\begin{figure}
\centering
\includegraphics[width=0.5\textwidth]{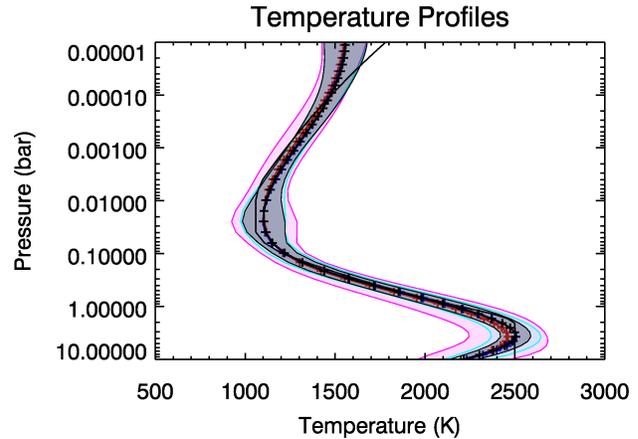}
\caption{As Figure~\ref{hotjuptempcompare}, but attempts have been made to correct for the baseline offsets between the NIRSpec and MIRI instruments. The improvement in the quality of the retrieval for an offset of order 3$\times$ the photon noise is significant, and a smaller improvement can also be seen for smaller offsets. \label{hotjuptempcorrected}}
\end{figure}

For the sun-like star, we calculate the spot contrast by simply assuming a star temperature of 5800 K and a spot temperature of 5000 K. A spot temperature 800 K cooler than the average is comparable to the best-fit spot temperature found for HD 189733b by \citet{pont13}. Since the shortest wavelength of NIRSpec occurs longward of the main spectral features in solar type stars, we assume that the star and spots can be represented by black bodies. The overestimation of the squared planet:star radius ratio is then caculated using the formula from \citet{pont13}:

\begin{equation}
1/{\alpha}=\frac{fB_{5000}+(1-f)B_{5800}}{B_{5800}}
\end{equation}

where the squared radius ratio is overestimated by a factor $\alpha$ if the fractional coverage of spots with flux $B_{5000}$ is $f$. To simulate this effect, we multiply the synthetic spectrum by $\alpha$. The presence of star spots will have a wavelength-dependent effect on spectra from the NIRSpec instrument, since the spot contrast increases rapidly towards the visible, but for MIRI we treat the star spot effect as uniform across the wavelength range for the sun-like star.

For the M dwarf case, we can no longer approximate the stellar disc and spot spectra as black bodies. This is because at cooler temperatures infrared absorption features, especially of H$_2$O, become more apparent in the spectra; since H$_2$O absorption is increased in the cooler spots, increased spot coverage could potentially mimic H$_2$O absorption in the planet's atmosphere in the case of unocculted spots. We therefore use PHOENIX model spectra at 3000 K for the stellar disc and 2300 K (the coolest available model) for the spot, and calculate the spot contrast as follows:

\begin{equation}
1/{\alpha}=\frac{fI_{spot}+(1-f)I_{star}}{I_{star}}
\end{equation}

At wavelengths longer than 5 microns we extrapolate these spectra to the closest black body curve. PHOENIX models at the same temperatures are also used for GJ 1214, with the appropriate metallicity and log($g$). 

One of the biggest problems likely to arise from star spots is spot coverage varying in the time between NIRSpec and MIRI observations of the same object. To simulate the effect of this, spots are applied at random to NIRSpec or MIRI observations, or both, or neither. This allows us to explore not only the effect of spots within a single observation, but also how activity variation can affect stitching of observations. For the M dwarf stars, which we expect to be more active, we test a spot coverage of 10\%, which is likely to represent the worst-case scenario. For the Sun-like star, the maximum spot coverage considered is 3\%, as in general these stars are less active than M dwarf stars. However, stellar activity levels for an individual star may vary significantly from the average for a given spectral type, so an activity study of interesting planet host stars before the \textit{JWST} launch would be extremely useful. 

\subsection{Offset Corrections}
\label{corrections}
\begin{figure*}
\centering
\includegraphics[width=0.85\textwidth]{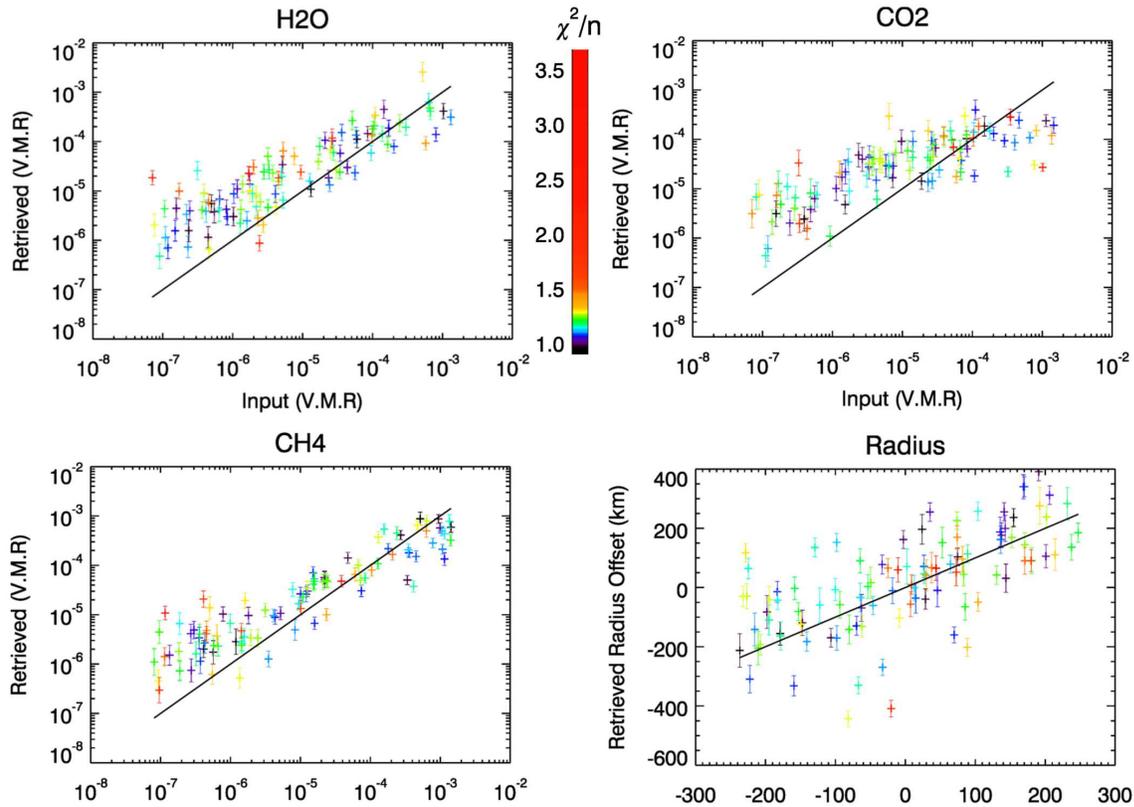}
\caption{100 retrievals of gas volume mixing ratio (VMR) and 10 bar radius for the hot Jupiter in primary transit. A scale factor on the input temperature profile is also retrieved. Colours correspond to reduced $\chi^2$ ($\chi^2$ divided by $n$, the number of measurements-number of free model parameters). The \textit{a priori} value for each gas is 10 ppmv.\label{hotj_primary_tempret}}
\end{figure*}

\begin{figure*}
\centering
\includegraphics[width=0.85\textwidth]{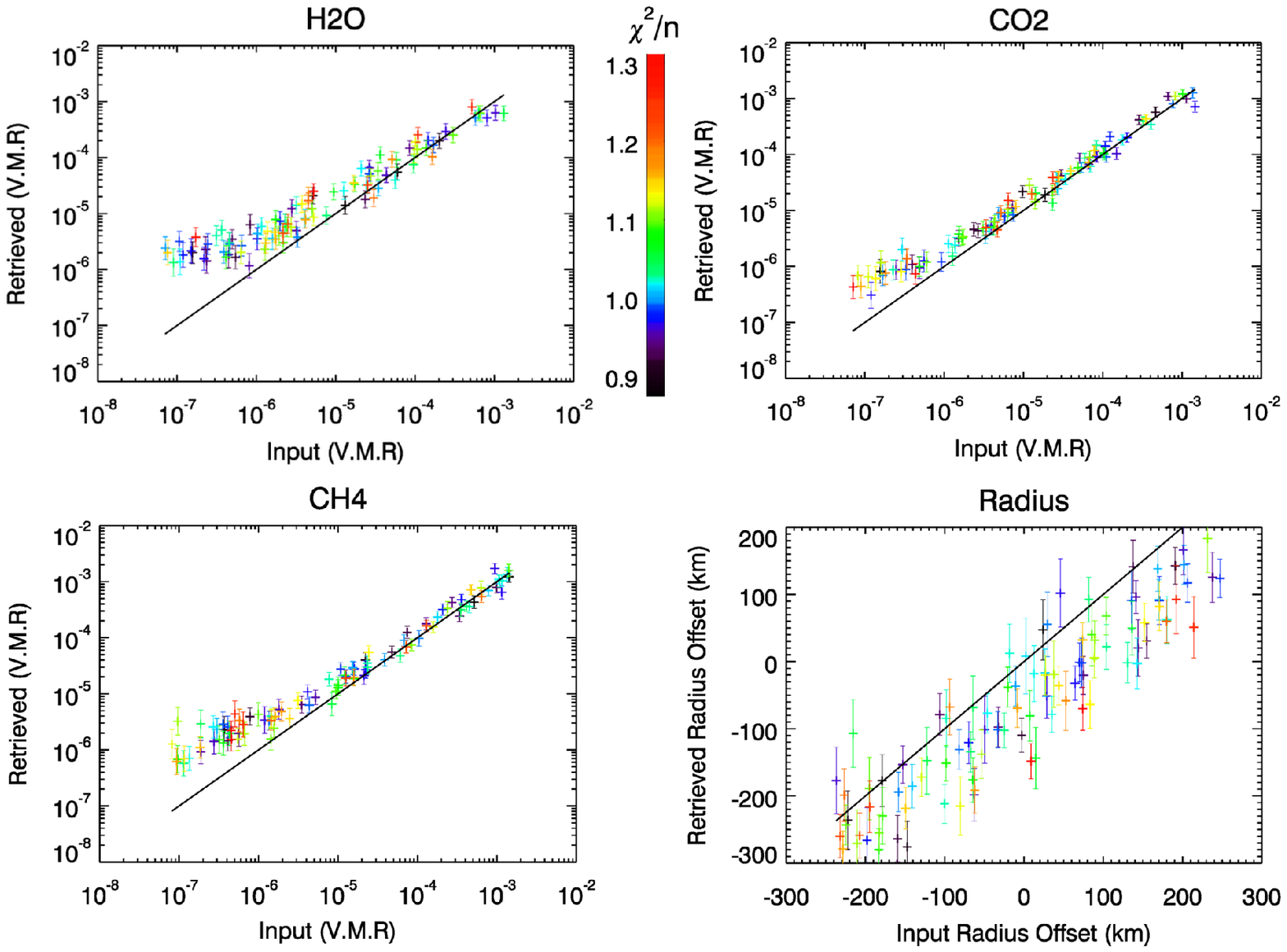}
\caption{As Figure~\ref{hotj_primary_tempret}, but this time the T-p profile is fixed to the known input. \label{hotj_primary}}
\end{figure*}
As well as exploring the effect of instrumental offsets and star spots, we also begin to investigate how these effects might be corrected for. Since NIRSpec and MIRI overlap over a small range of wavelengths, albeit in a part of the MIRI spectrum where the throughput is low, it may be possible to tell that an offset is present between the spectra measured by different instruments. If the shift is uniform, as assumed in the cases we test, it may also be possible to correct for it. 

To test this, we use NIRSpec and MIRI spectra with a uniform offset applied. We average the transit depth for each instrument in the overlapping wavelength region and calculate the difference. We then correct the MIRI spectrum relative to the NIRSpec spectrum according to the measured offset, and see if this improves the retrieval accuracy over the uncorrected case. 

We have not considered here instrumental systematics that result in a varying baseline across a given instrument. It is to be hoped that careful detector calibration and a detailed understanding of detector behaviour would allow these effects to be mitigated.

Without monitoring the host star, as has been done for HD 189733b in the work of \citet{pont13}, it will not be possible to differentiate between an offset due to instrumental effects or due to stellar activity in primary transit observations. Therefore, the correction applied only assumes a wavelength-invariant shift of the transit depth seen in one instrument compared to the other. This will not account for the non-grey effect on starspots on the NIRSpec spectrum. 

\section{Code}
\label{code}

We use the NEMESIS radiative transfer and retrieval code, which was originally developed for solar system planets and has since been expanded for exoplanet modelling applications. For details of its use for real exoplanet observations, see \citet{lee12,barstow13b,lee13a,barstow14} and \citet{lee13}. 

NEMESIS incorporates a fast, correlated-k based radiative transfer model with an optimal estimation retrieval scheme. Whilst optimal estimation is limited by its Gaussian treatment of errors, in cases where observed spectra contain sufficient information such schemes perform as well as more rigorous, but slower, algorithms such as Differential Evolution/Markov-Chain Monte Carlo \citep{line13}. Correlated-k \citep{goodyyung,lacis91} is an approximation applied to gas absorption features within the radiative transfer calculation, and allows gas absorption ($k$) coefficients to be calculated offline and pre-tabulated, increasing the speed of the model run. K-tables can be used without significant loss of accuracy provided that the relative strengths of absorption lines are well-correlated between adjacent layers of the model atmosphere - e.g. the strongest line at the bottom of the atmosphere is also the strongest line in the next layer up. This approach has been extensively used to successfully model various atmospheres within the solar system (e.g. \citealt{irwin04,fletcher11,barstow12}). 

We include collision-induced absorption due to H$_2$ and He within the model atmospheres, as well as absorption line data for H$_2$O, CO$_2$, CO and CH$_4$. The sources for these data are listed in Table~\ref{gasdata}. In the Earth model we include data for H$_2$O, CO$_2$, CO, O$_2$, O$_3$, CH$_4$ and N$_2$O, all from the HITRAN 2008 database \citep{roth09} . 

\section{Results}
\label{results}

\subsection{Hot Jupiter}
The secondary transit case with \textit{JWST} is the most straightforward to deal with, because star spots are not likely to strongly affect these results. The only stitching problem we need to consider is instrumental offsets between NIRSpec and MIRI, so we discuss secondary transit first for each of the planetary types listed in Table~\ref{planet_properties}. The advantage of secondary transit observations is that the thermal flux emerging from the planet over a range of pressures is measured, and so the temperature structure of the atmosphere can be inferred. 
\subsubsection{Secondary eclipse}
The first test case is a clear-atmosphere hot Jupiter orbiting a sun-like star at 250 pc. The signal-to-noise ratio for this case is sufficient for spectral features to be detected (Figure~\ref{hotjup_spectra}), so assuming no systematic instrumental offset a successful retrieval of temperature and H$_2$O, CO$_2$ and CH$_4$ VMRs is possible provided each species has a volume mixing ratio of at least 0.5 ppmv, as found by \citet{barstow13} for \textit{EChO} (Table~\ref{results_occultation}). A good retrieval of CO is more difficult as this molecule has fewer spectral features in the wavelength region considered. However, if there is a systematic offset between the NIRSpec and MIRI instruments the quality of the retrieval is seriously degraded (Figure~\ref{hotjuptempcompare}). This is because the two instruments probe different gases at different pressures within the atmosphere (see Figure~\ref{tempcov} for sensitivity at different pressures as a function of wavelength), and so the offset produces artefacts in the retrieved temperature profile and gas abundances by increasing or decreasing the flux apparently emerging from the upper relative to the lower atmosphere, or vice versa. An offset on the order of the photon noise has a noticeable but fairly small effect on the typical accuracy of the retrieval, especially at higher pressures, and a larger offset of around 3$\times$ the photon noise seriously reduces the accuracy everywhere, again with the largest effect in the deep atmosphere as the sensitivity is lower at higher pressures. 

However, reassuringly, implementing the simple correction for the shift we describe in Section~\ref{corrections} enormously improves the quality of the retrieval (Figure~\ref{hotjuptempcorrected}). Whilst the spread of retrieved temperature profiles is still larger for offset cases, because only the relative offset between the two instruments is corrected for, this type of correction should allow us to have confidence in combined NIRSpec and MIRI observations. 

\subsubsection{Primary transit}

We also test the accuracy of retrieved gaseous abundances in primary transit for the model planets. As discussed in \citet{barstow13}, retrieving a continuous temperature profile from a primary transit spectrum is not possible, since the deep atmosphere is not probed and therefore there is insufficient information to fully constrain the T-p profile. Another issue, mentioned in \citet{barstow13b}, is that the planetary radius assumed at a given pressure level has a significant effect on primary transit retrievals. Since the planet radius is derived from the white light transit, it is dependent on the atmospheric properties and so must be included as a free parameter in a primary transit retrieval. In this case, we retrieve the radius at the 10-bar pressure level in the atmosphere, which is the base of our model atmosphere. The atmosphere is assumed to be opaque at all wavelengths for pressures greater than this (in practice, the atmosphere is opaque at all pressures $>$1 bar, see Figure~\ref{tempcov} for sensitivity to different pressures in primary transit). Increasing the 10-bar radius therefore has two effects; firstly, it increases the total absorbing area of the planet, increasing the transit radius in a grey sense across all wavelengths; secondly, it decreases the gravitational acceleration at a given pressure (as this is proportional to $1/r^2$) thus increasing the atmospheric scale height. These two effects result in significant and complex degeneracy between the 10-bar radius and other parameters. For example, increasing the H$_2$O abundance in the model would increase the size of H$_2$O absorption features in the spectrum, which might trade off with a smaller 10-bar radius to keep the transit depth constant; however, increasing the H$_2$O abundance will also increase the mean molecular weight of the atmosphere, decreasing the scale height, and this might then trade off with a larger 10-bar radius which has the effect of increasing the scale height.

\begin{figure}
\centering
\includegraphics[width=0.5\textwidth]{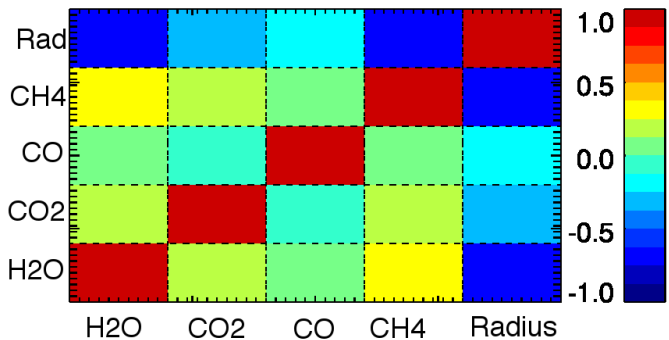}
\caption{The correlation matrix for a single hot Jupiter primary transit retrieval (temperature is not retrieved in this case). There are significant negative correlations between the 10-bar radius offset and the gas abundances. \label{hotj_cor}}
\end{figure}

Although a continuous T-p profile retrieval is not possible in primary transit, the spectral shape is not completely independent of temperature structure. For the hot Jupiter case, where the secondary transit is easily observable, we will have some prior information on the expected shape of the T-p profile. However, the terminator T-p profile is very likely to be cooler than the dayside profile, and won't necessarily display exactly the same vertical shape, as observed for the hot Jupiter WASP-43b by \citet{stevenson14b}. Therefore, we also test the effect of using a generic adiabatic T-p profile of an appropriate effective temperature, and including a scaling factor on this profile in the retrieval. This allows some constraint to be placed on the stratospheric temperature, and should ensure an accurate retrieval of gaseous abundances even when the T-p profile is unknown.

In Figure~\ref{hotj_primary_tempret} we show the retrieved versus input results for the case where the temperature is scaled; in Figure~\ref{hotj_primary} we show the same results for the case where input temperature profile is used. The scatter on the results in Figure~\ref{hotj_primary_tempret} is much larger than seen in Figure~\ref{hotj_primary}, as the addition of the temperature parameter increases retrieval degeneracy. To resolve this difficulty for a hot enough target, obtaining a full phase curve as in \citet{stevenson14b} would provide more direct information about the terminator temperature. It is clear from Figure~\ref{hotj_primary} that, even where the input temperature is used, there is some degeneracy between the retrieved gas abundances and the 10-bar radius. There is significant negative correlation between 10-bar radius and the gas abundances, resulting in a systematic underestimation of the 10-bar radius and overestimation of the gas abundances. We show the correlation matrix from the retrieval in Figure~\ref{hotj_cor}; this is the normalised error covariance matrix for the atmospheric state vector, showing correlated errors between different variables. Significant correlations, as seen here, mean that the retrieval is degenerate. This can be rectified by observing multiple transits to reduce the noise on the spectrum - observing 10 transits provides tighter constraints and reduces the degeneracy, removing these systematic under- and over-estimations. The retrieval accuracies for a single transit are listed in Table~\ref{results_transmission}.
\begin{figure*}
\centering
\includegraphics[width=0.85\textwidth]{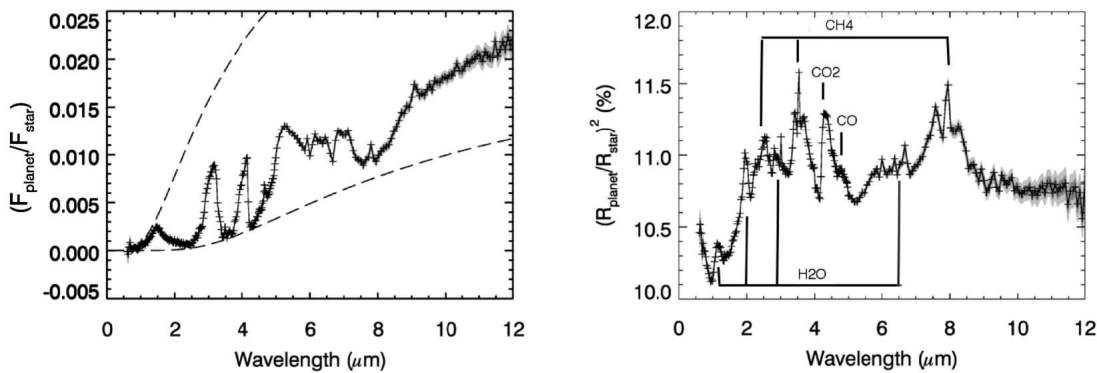}
\caption{As Figure~\ref{hotjup_spectra} but for single secondary/primary transits of the hot Neptune orbiting an M dwarf. The input temperature profile is shown in Figure~\ref{hotnep}. Dashed lines on the left indicate the blackbody flux ratios at 1500 K, top, and 750 K, bottom. We also show the positions of the main absorption bands on the right-hand plot as they are easy to distinguish for this case.\label{hotnep_spectra}}
\end{figure*}

As for the secondary transit case, instrumental systematics may need to be accounted for, and for primary transit there is also the possibility of variable stellar activity causing problems. We test this by adding an unocculted star spot offset, as described in Section~\ref{noise}, at random to some NIRSpec and some MIRI spectra within the test set of 500 models. This means that some combined spectra are spot-free, some have the same spot offset in both NIRSpec and MIRI sections, and some spectra have a spot offset in either the NIRSpec or MIRI section but not in both. 

For a sun-like star, we find that even for a reasonably high spot coverage (3\%) the quality of the retrieval is not significantly decreased. This is because the peak in star/spot contrast occurs at wavelengths shorter than \textit{JWST}'s range, so especially for the MIRI instrument spots have very little effect. Baseline offsets between instruments can also largely be accounted for. 
\begin{figure*}
\centering
\includegraphics[width=0.85\textwidth]{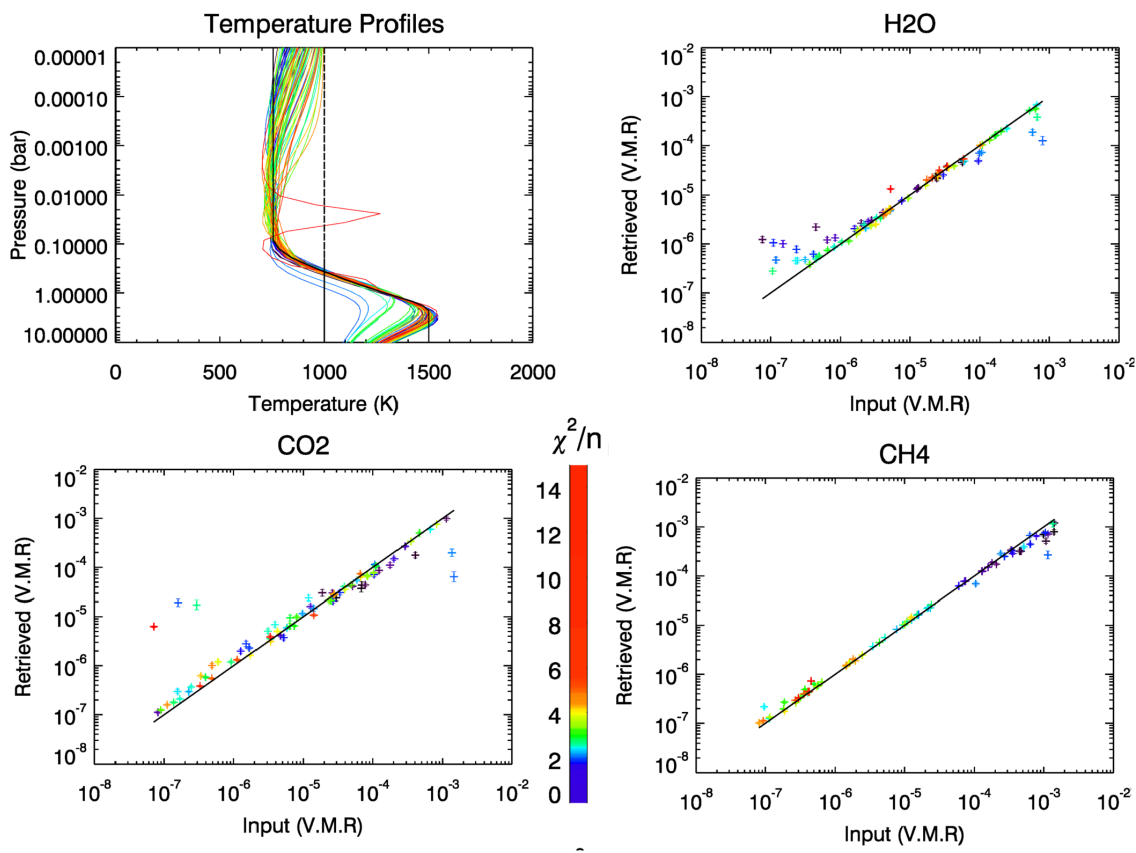}
\caption{100 eclipse retrievals of temperature and gas abundances for a hot Neptune orbiting an M dwarf at 15 pc distance. Colours correspond to $\chi^2/n$. As for the hot Jupiter, temperature and H$_2$O, CO$_2$ and CH$_4$ VMRs are retrieved accurately in most cases, apart from one retrieval with a high reduced $\chi^2$ which has low abundances of all gases so fewer detectable features.\label{hotnep}}
\end{figure*}

For a cloud-free, well-mixed atmosphere, the gas abundance information retrieved from primary transit is very similar to that retrieved from secondary transit. However, primary transit measurements are still necessary, despite the added complications from stellar activity, because a) the atmosphere at the terminator may be different from the dayside, and b) if the atmosphere is cloudy or not well-mixed then the information content of the two geometries is very different. Generally, primary transit spectra probe higher regions of the atmosphere due to the longer path length in limb geometry, so for a case where photochemistry alters the composition of the atmosphere at higher altitudes there might be a wealth of extra information to be gained, and of course primary transit spectra are excellent indicators of the presence of clouds.

\begin{table}
\begin{tabular}[c]{|c|c|c|c|}
\hline
Gas & Hot Jupiter & Hot Neptune\\
\hline
H$_2$O & 3$\times$ ($>$0.1 ppmv) & 1.5$\times$ ($>$1 ppmv) \\
CO$_2$ & 2$\times$ ($>$0.5 ppmv) & 3$\times$ ($>$10 ppmv)\\
CO &  10$\times$ ($>$10 ppmv)  & 5$\times$ ($>$10 ppmv) \\
CH$_4$ & 3$\times$ ($>$0.1 ppmv) & 2$\times$ ($>$0.1 ppmv)\\
\hline
\end{tabular}
\caption{Achievable precision in gas abundance retrievals from a single observation for the hot Jupiter and hot Neptune in eclipse. These precisions are only valid for certain concentrations, which ranges are given in parentheses for each case. As the gas abundances vary over several orders of magnitude, the precisions are given as multiples of the abundance; for example, the H$_2$O VMR can be retrieved to within a factor 3 of the true value for a hot Jupiter, provided there is more than 0.1 ppmv present. The values quoted take into account the errors due to retrieval degeneracy, which is the dominant source of error, by calculating the precision based on the 2$\sigma$ deviation from the true value over all retrievals. These are calculated assuming no systematic offsets. The hot Neptune case has a higher signal to noise as the system is significantly closer and the planet:star radius ratio is larger.\label{results_occultation}}
\end{table}

\begin{table}
\begin{tabular}[c]{|c|c|c|c|}
\hline
Gas & Hot Jupiter & Hot Neptune & GJ 1214b\\
\hline
H$_2$O & 3$\times$ ($>$10 ppmv) & 1.5$\times$ ($>$0.1 ppmv) &  5$\times$ ($>$50 ppmv)\\
CO$_2$ & 2$\times$ ($>$1 ppmv) & 1.5$\times$ ($>$0.1 ppmv) & 5$\times$ ($>$50 ppmv)\\
CO & Not detectable & 1.5$\times$ ($>$10 ppmv) & N/A \\
CH$_4$ & 3$\times$ ($>$5 ppmv) & 1.5$\times$ ($>$0.1 ppmv) & 5$\times$ ($>$50 ppmv)\\

\hline
\end{tabular}
\caption{Information available from a single observation for the hot Jupiter, hot Neptune and GJ 1214b in transmission. Perfect knowledge of the temperature profile (and cloud top altitude in the case of GJ 1214b) has been assumed. Otherwise everything is as Table~\ref{results_occultation}.\label{results_transmission}}
\end{table}

\subsection{Hot Neptune}
\begin{figure*}
\centering
\includegraphics[width=0.85\textwidth]{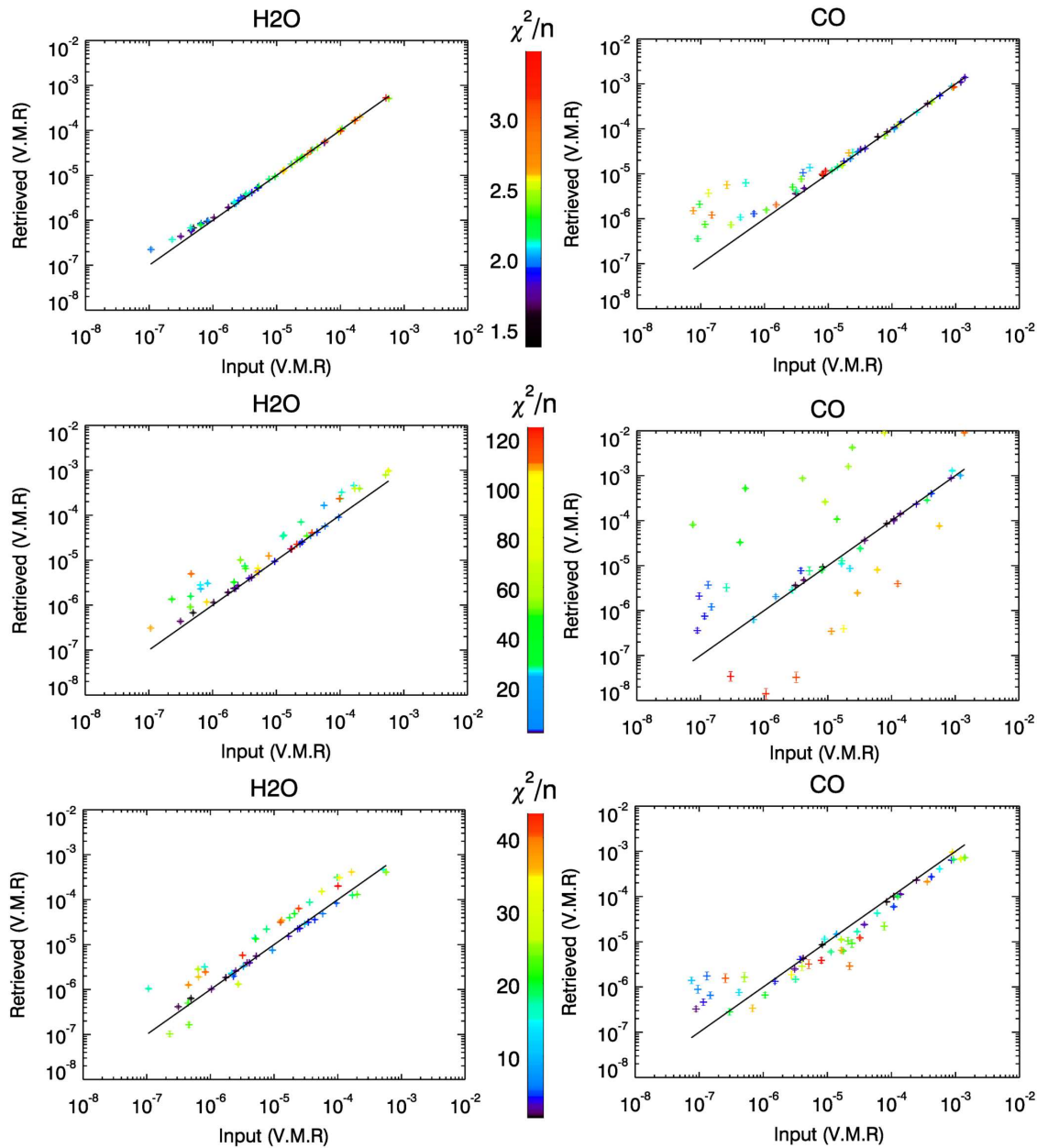}
\caption{The effect of star spots on a hot Neptune primary transit observation. Top row: spot-free case. Middle row: 10\% spot coverage in some observations. Bottom row: 10\% spot coverage but shift between NIRSpec and MIRI corrected. \label{hot_nep_primary_spots}}
\end{figure*}
\subsubsection{Secondary eclipse}
As in \citet{barstow13} we consider a hot Neptune in a close orbit around a cool M dwarf. In this work, the system is placed at 15 pc, almost three times further away than the equivalent in \citet{barstow13}, to avoid the risk of saturation. Spectra are shown in Figure~\ref{hotnep_spectra}. With a single secondary transit observed with each instrument a good constraint can be obtained on H$_2$O, CO$_2$ and CH$_4$ abundances, as well as the tropospheric temperature profile (Figure~\ref{hotnep}). Again, a retrieval of CO is more difficult than retrievals of other species due to the smaller number of CO lines. 

We also find that introducing an offset between NIRSpec and MIRI has a similar effect for the hot Neptune as we find with the hot Jupiter. Assuming there is sufficient overlap between the two instruments, applying a shift to correct for the offset means that the quality of the retrieval is not noticeably degraded. For secondary transit, with sufficiently high SNR, systematic effects such as this should not result in problems. 

\subsubsection{Primary transit}

The effect of star spots is most critical for planets orbiting M dwarfs. These cooler stars generally have higher activity, and because of their lower temperatures their spectra contain some molecular absorption features, particularly due to water vapour. These spectral features will have different shapes within the cooler spot regions compared with the rest of the disc, and therefore unocculted spots could not only lead to an overestimation of the radius ratio at shorter wavelengths, but also inaccurate retrieved values of water vapour abundance. 

We present in Figure~\ref{hot_nep_primary_spots} the effect of 10\% spot coverage on the retrieval of H$_2$O and CO abundances from a single primary transit of the hot Neptune. The retrievals for the spot-free, no-offset case (row 1) show that H$_2$O can be retrieved to very high precision, with CO well retrieved as well for abundances down to 10 ppmv. However, if instead the star has 10\% out-of-transit-chord spot coverage for some of the observations, it is clear that in many cases the H$_2$O abundance is overestimated by up to an order of magnitude (row 2). This is due to the differing water vapour signatures in the stellar and spot spectra mimicking increased water vapour absorption in the planet's atmosphere. Correcting for the offset between NIRSpec and MIRI (row 3) doesn't solve this problem; however, it does improve the CO retrieval from the uncorrected spot case. We can see from this that for CO the offset between the two instruments is the most important factor, whereas this is not the case for H$_2$O. This result shows that stellar activity may be the limiting factor for accurate retrievals of \textit{JWST} observations for planets around active stars. 

The situation presented above is a worst-case scenario, since it may be possible to estimate the effect of star spots by observing spot crossing events during transits and exploiting \textit{JWST}'s wavelength coverage. \citet{fraine14} use evidence of spot crossings in simultaneous Kepler and Spitzer lightcurves of the warm Neptune HAT-P-11 b to place constraints on the temperature contrast of the star spots. Spot crossings are clearly observed in the Kepler lightcurves but not at all in the 4.5 $\upmu$m Spitzer data and only indirectly in the 3.6 $\upmu$m data, so a lower limit on the spot temperature can be obtained. In this case, even the coolest spots do not present a strong enough contrast to mimic the H$_2$O absorption seen in the planet's atmosphere. It will not be possible to obtain simultaneous observations of spot crossings over \textit{JWST}'s full wavelength range, but observation of several transits with both NIRSpec and MIRI could provide a statistical estimate of the average spot temperature, assuming that the planet's orbit crosses the spot latitudes on the star. If this is not the case, then scenarios like the one described above could still be problematic, although as close-in planets generally have misaligned orbits some spot crossings are likely.

\subsection{GJ 1214b}
\begin{figure}
\centering
\includegraphics[width=0.5\textwidth]{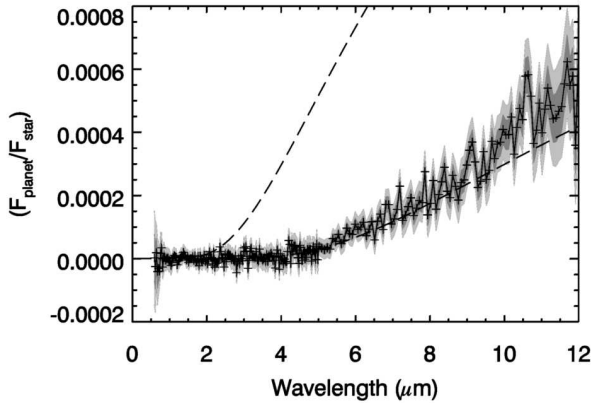}
\caption{As for Figure~\ref{hotjup_spectra}, this is a secondary transit spectrum for 15 eclipses of GJ 1214b. No spectral features are visible above the noise level. Dashed lines indicate the blackbody flux ratios at 800 K, top, and 450 K, bottom.\label{gj1214b_sec_spec}}
\end{figure}

\subsubsection{Secondary eclipse}
No observation of GJ 1214b in secondary transit has yet been possible, as current instruments are not sensitive enough to detect it since the planet is relatively cool. We model here what a secondary eclipse of GJ 1214b might be expected to look like (Figure~\ref{gj1214b_sec_spec}). Whilst the secondary eclipse is certainly detectable with \textit{JWST} and the broad shape of the spectrum is discernible, it is likely that the presence of clouds on the planet coupled with a low SNR will confuse interpretation of the secondary transit spectrum. This means that retrievals of gaseous abundances and a detailed retrieved temperature profile from secondary transit are unlikely. In addition, very little flux is observable coming from the planet in the NIRSpec range, so just using MIRI is likely to be adequate and therefore systematic offsets between the instruments are not a concern here. 

The most useful piece of information we could gain from a secondary transit of GJ 1214b is a constraint on the stratospheric temperature, since the MIRI weighting functions peak in the stratsophere. Allowing for likely variation between the dayside and the limb, this could be used as prior information for primary transit retrievals. It could help to break degeneracies between the atmospheric temperature, mean molecular weight and 10-bar radius parameter, all of which affect the scale height of the atmosphere. 

\begin{figure}
\centering
\includegraphics[width=0.5\textwidth]{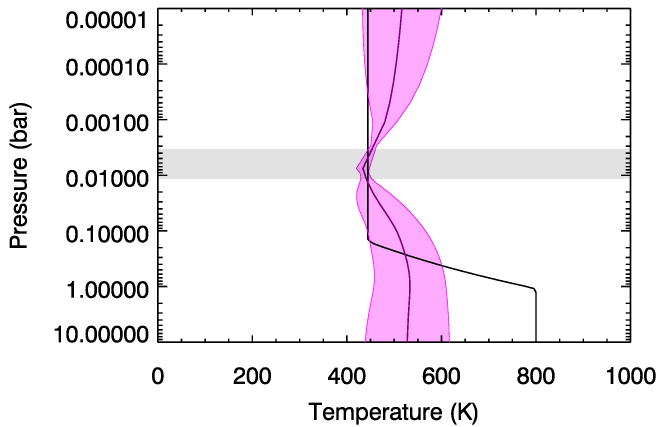}
\caption{An average of six temperature retrievals from 15 eclipses of GJ 1214b, starting from six different isothermal priors between 400 and 650 K. All retrievals converge on the correct temperature at around 0.01 bar, within the grey shaded area. The purple line is the average, the pink shading indicates the spread of different retrieved values and the solid black line is the input temperature profile.  \label{gj1214btempret}}
\end{figure}

For 15 eclipses of GJ 1214b, a secondary transit spectrum can be obtained with sufficiently high SNR to constrain the stratospheric temperature at 0.01 bar by fitting the near-blackbody radiance (Figure~\ref{gj1214btempret}). Whilst there is clearly not enough information in the spectrum to retrieve the full profile, due to the lack of spectral features, in primary transit regions of the atmosphere below this are not probed anyway, so the detailed temperature structure at lower altitudes is not relevant to primary transit retrievals. 

\subsubsection{Primary transit}

GJ 1214b has already been extensively observed in primary transit, but over the limited wavelength range available in space and from the ground (less than 5 $\upmu$m) since Spitzer lost its cryogenic cooling capability. GJ 1214b has posed a challenge for transit spectroscopy due to its extremely flat spectrum, which must be due to the presence of cloud or haze high in its atmosphere \citep{kreidberg14}. Because the spectrum as currently observed is featureless, no conclusions can be drawn about the composition of the atmosphere. Reducing the noise on the spectrum and increasing the wavelength coverage should help to solve this problem. 

\begin{figure*}
\centering
\includegraphics[width=0.85\textwidth]{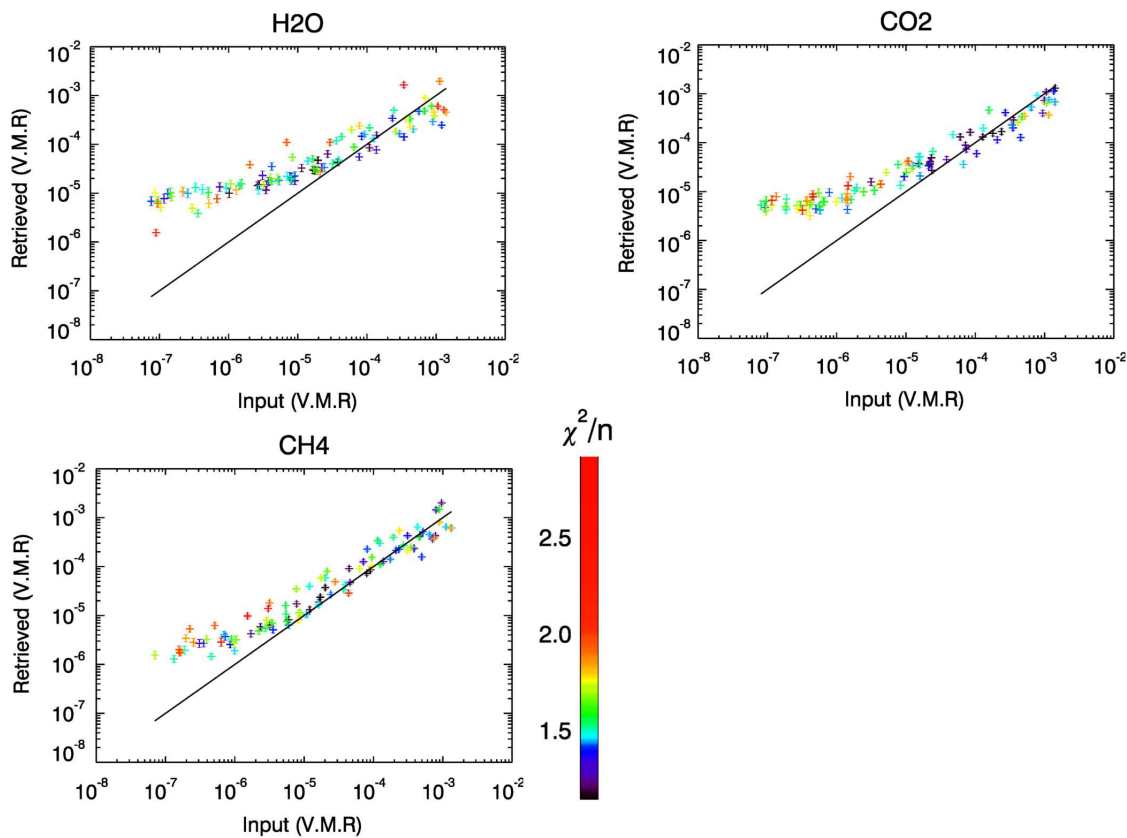}
\caption{100 retrievals for a single transit of GJ 1214b observed with JWST. Gases are only well retrieved if the abundances are above a 50 ppmv threshold, when the features can be distinguished above the broad signature of the cloud. Gas priors are 100 ppmv. \label{gj1214b_primary}}
\end{figure*}

\begin{figure}
\centering
\includegraphics[width=0.5\textwidth]{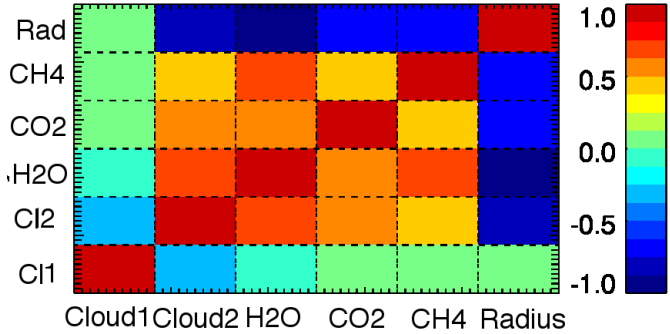}
\caption{The correlation matrix for a single GJ 1214b primary transit retrieval (temperature and cloud top pressure are not retrieved in this case). There are significant negative correlations between the 10-bar radius offset and the gas abundances. \label{gj1214b_cor}}
\end{figure}

We present retrievals (temperature and cloud top pressure are fixed at the correct input values) for a single transit of GJ 1214b, assuming cloud properties as modelled in \citet{barstow13b}. Values for H$_2$O, CO$_2$ and CH$_4$ can be accurately retrieved to within a factor of 3 if the abundance of any of these gases is above 10 ppmv (Figure~\ref{gj1214b_primary}), because only if they are sufficiently abundant can their spectral features be seen above the cloud deck (1 mbar in this case). The cloud optical depths and radius offset are not well retrieved, due to high correlations between variables (Figure~\ref{gj1214b_cor}). For a detailed discussion of degeneracies between cloud top pressure and temperature in primary transit observations, see \citet{barstow13b}.

\begin{figure}
\centering
\includegraphics[width=0.5\textwidth]{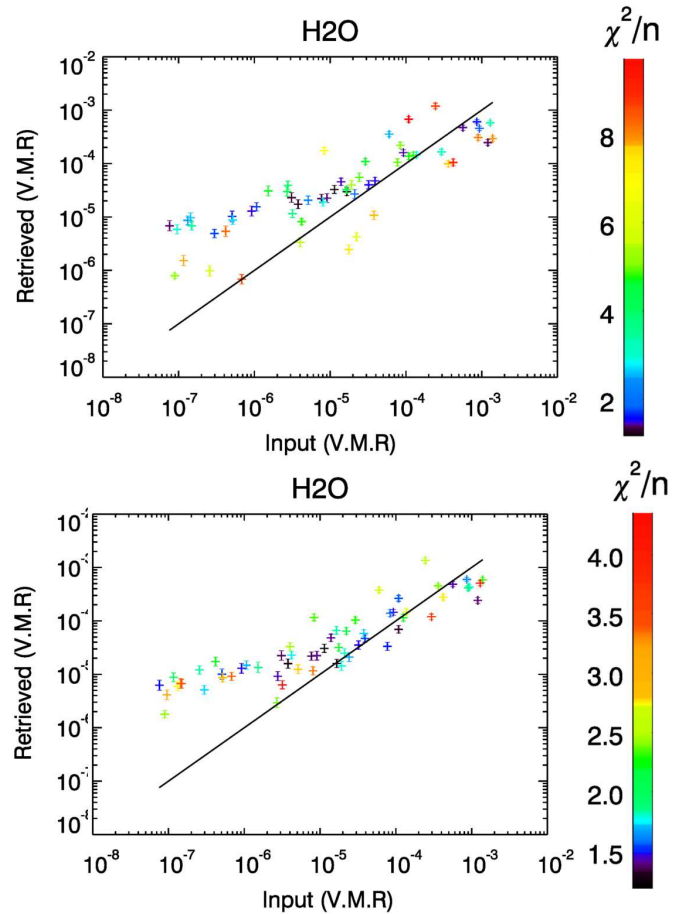}
\caption{The effect of 10\% out-of-transit star spot coverage on the retrieval of water vapour for a GJ 1214b-like planet. As for the hot Neptune, the presence of unocculted starspots can lead to an overestimation of water vapour that is difficult to correct for.\label{gj1214b_spots}}
\end{figure}

As for the hot Neptune case, star spots introduce further difficulties with gas retrievals for GJ 1214b, although the effect is less noticeable due to the relatively poor quality of the retrieval for the spot free case. If the spot coverage is up to 10\%, water vapour abundance can be overestimated by as much as an order of magnitude (Figure~\ref{gj1214b_spots}), and uncorrected offsets between NIRSpec and MIRI also increase the scatter on all retrieved gas abundances. 

Whilst \textit{JWST} represents an enormous improvement on current instrumentation for a planet like GJ 1214b, atmospheric retrievals are still not straightforward due to the cloudiness of the atmosphere. The effect of the clouds is to obscure gaseous absorption features, thus reducing the information content of the spectrum and making it more difficult to break degeneracies in retrieval. This can be improved upon further by combining the results of multiple observations to increase the SNR of any atmospheric features that are visible, but ultimately the extent of the cloud coverage is the limiting factor for transit spectroscopy of this kind of planet.

\begin{figure*}
\centering
\includegraphics[width=0.85\textwidth]{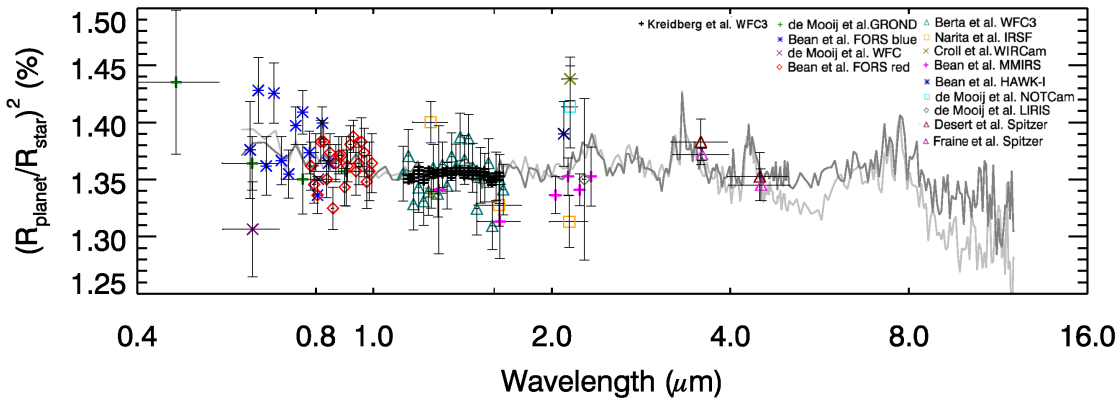}
\caption{Primary transit spectra based on two different atmospheric models for GJ 1214b (15 transits). Both are compatible with current data, but they diverge at wavelengths that will be sampled for the first time with \textit{JWST}. Existing data are as shown in \citet{barstow13b}, with the data from \citet{kreidberg14} and \citet{fraine13} added. The dark/light grey lines show models with the cloud top at 0.1/0.01 mbar. \label{gj1214bcloudtop}}
\end{figure*}

\begin{figure*}
\centering
\includegraphics[width=0.8\textwidth]{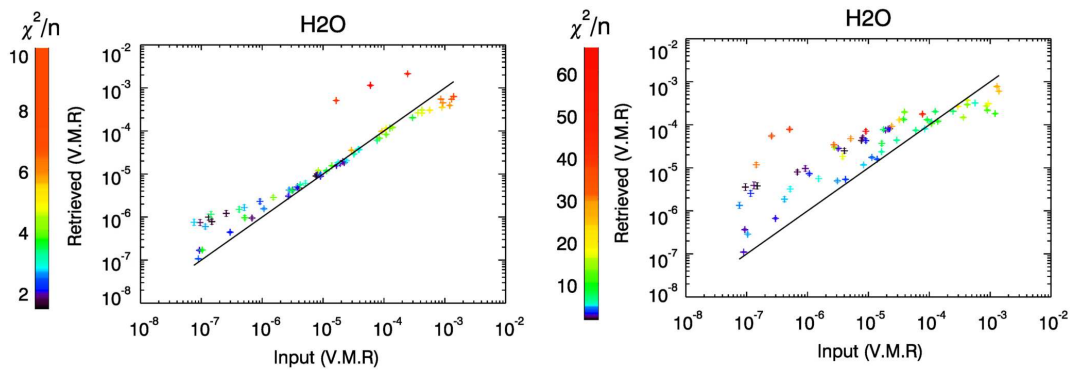}
\caption{Retrieval of H$_2$O abundance from 15 transits of GJ 1214b if the cloud top pressure used in the retrieval is the same as the input (1 mbar, left) and if it is higher (0.1 mbar, right). The $\chi^2/n$ values where the incorrect cloud top pressure is used are higher, indicating that the fit is not as good and demonstrating sensitivity to cloud top pressure.\label{cloudalt}}
\end{figure*}

Despite these degeneracies, some constraint on the cloud can be obtained with \textit{JWST}. The current data from \textit{HST}/WFC3 \citep{kreidberg14} allow an upper limit to be placed on the cloud top pressure, but a range of cloudy atmosphere models with the cloud at lower pressures are still permitted. In Figure~\ref{gj1214bcloudtop}, it is clear that over the wavelength range probed by \textit{JWST}/NIRSpec+MIRI the cloud top pressure degeneracy starts to be broken. In this figure, we show noisy synthetic spectra based on two different atmospheric models, one with a cloud top at 0.01 mbar and one with a cloud top at 0.1 mbar. Both models produce an equally good fit to the existing data, but have rather different properties. To compensate for the higher cloud top, the 0.01 mbar model has a smaller cloud optical depth, which results in the extinction at longer wavelengths being smaller than for the 0.1 mbar model which produces a much flatter spectrum.  In Figure~\ref{cloudalt}, we demonstrate sensitivity to cloud top pressure by comparing retrievals of H$_2$O abundance from 15 transits of GJ 1214b, for the case where the input cloud top pressure is used in the retrieval versus the case where a higher cloud top pressure is used. The spectral fit is consistently worse if the wrong cloud top pressure is used, again demonstrating that we can use \textit{JWST} observations to place some constraint on cloud top pressure for GJ 1214b.

\subsection{Earth}

We perform a final test in primary transit for an Earth analog planet in orbit around an M dwarf at 10 pc distance. This is a very difficult target to observe even with the sensitivity of a telescope like \textit{JWST}, but if a close enough system should be discovered it is necessary to think about what we might be able to see. 

\begin{figure*}
\centering
\includegraphics[width=0.85\textwidth]{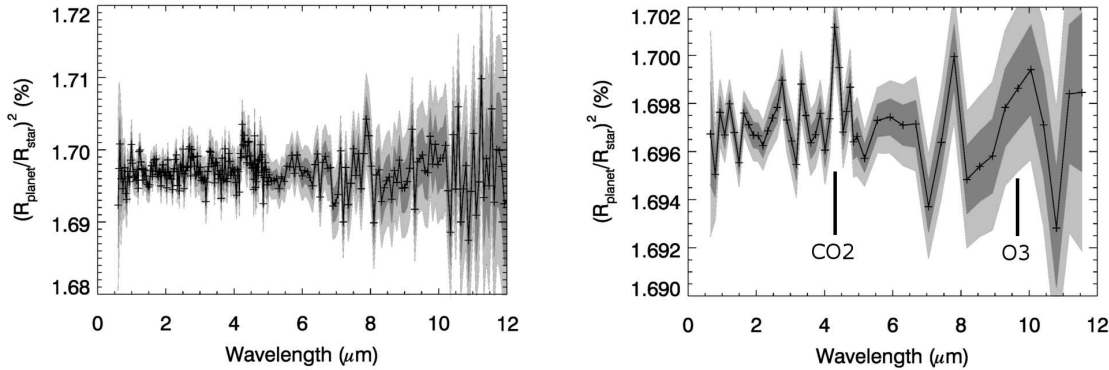}
\caption{As Figure~\ref{hotjup_spectra} but for 30 primary transits of an Earth orbiting an M dwarf. The spectrum on the right is binned up by a factor 5 to make the ozone band at 9.6 microns more obvious. The CO$_2$ band at 4.3 microns is also clearly visible.\label{earth_spectra}}
\end{figure*}

Ultimately, we would like to be able to detect gases that might indicate an Earthlike environment, or even life. Ozone is a key infrared absorber that is chemically linked to diatomic oxygen, so it would be extremely interesting if we could detect this in the atmosphere of a temperate, Earth-size planet. We simulate a spectrum for 30 transits of an Earth in orbit around an M dwarf, which would span of order one year as each orbit would take approximately 10 days. We use the Earth atmosphere model from \citet{irwin14}. Although the resultant primary transit spectrum is clearly very noisy indeed, it is possible to pick out the 4.3-$\upmu$m CO$_2$ band and the 9.6-$\upmu$m O$_3$ band; these can be seen more clearly in the binned spectrum (Figure~\ref{earth_spectra}). The spectrum is too noisy for an accurate retrieval of the gas abundances, but the presence of ozone can be detected using a retrieval technique; even when the ozone prior is set to very low values, the retrieved value is similar to the true abundance. No offsets or star spot corrections were applied to this spectrum, but even in the presence of these systematics we would expect this feature to remain detectable as a) the feature sits in the middle of the MIRI wavelength range, so should be relatively unaffected by instrumental baseline offsets and b) the effect of star spots at these wavelengths is smaller and relatively grey. We would also not expect to see an O$_3$ feature within a stellar spectrum; H$_2$O features within the stellar and spot spectra are the main cause of the problems retrieving H$_2$O in the hot Neptune case presented here.

Of course, the photochemical environment of an Earth-like planet in a close orbit around an M5 star is very different from the environment around the sun. \citet{grenfell13} find that O$_3$ would still be produced around cooler stars, but the mechanism is likely to be different and this may result in differences in the vertical profile. It could therefore conceivably be the case that O$_3$ is produced, but not detected because it is deeper in the atmosphere and the signature is obscured by clouds.

\section{Discussion}
\label{discussion}
\textit{JWST} will clearly be a powerful observatory for transit spectroscopy, and undoubtedly for other methods of exoplanet characterisation not discussed here. However, there is a significant amount of work to be done in order to guarantee that we can fully exploit its potential. \textit{JWST}'s lifetime is limited by the amount of propellant it can carry, and so it is especially crucial that multiple observations of small, low SNR targets are planned well in advance. We here identify several areas for necessary future work.

\subsection{Detector systematics}
We have highlighted the issues that can arise if there are baseline offsets between instruments for which we are stitching together observations, and discussed ways in which these might be corrected. The effect of these offsets before correction makes it clear that detector systematics are likely to be important in transit spectroscopy with \textit{JWST}, and a baseline offset is only one simple example. Infrared detectors such as those in the \textit{Spitzer} IRAC instrument can display intrapixel gain variability (seen by e.g. \citealt{knutson08} for HD 209458b, \citealt{anderson11} for WASP-17b) and other more complex systematic effects that require more sophisticated techniques to correct for than those discussed in this work. Methods such as independent component analysis (e.g. \citealt{waldmann14}), Gaussian Processes \citep{gibson14} and pixel mapping have all been used to correct for these more complex effects when reducing the observed data. However, for \textit{JWST} we have the advantage of studying the detector systematic behaviour prior to launch with a view to using them for exoplanet spectroscopy - a luxury that was unavailable for \textit{HST} and \textit{Spitzer}. This, combined with the continuing development of data reduction techniques, should ensure that detector systematics are less of an obstacle for \textit{JWST}. 

\subsection{Starspot monitoring and activity characterisation}
It is clear that, particularly for cooler stars, the effects of stellar activity and especially unocculted spots must be accounted for in transit spectra. At short wavelengths, unocculted spots can mimic a larger planet:star radius ratio and also a larger water vapour abundance. This could potentially result in false positive detections of cloud and water vapour, as well as inaccurate retrievals of other quantities. Unfortunately, it is difficult to blindly separate the effects of stellar activity from genuine atmospheric features within transit spectra. 

Attempts have been made to overcome this difficulty for recent observations of HD 189733b. \citet{pont13} correct for the effects of starspots by monitoring the flux of the host star HD 189733 over a long period of time using the Automated Patrol Telescope at Siding Spring Observatory, which provides an indication of the varying spot coverage. This time series is used to estimate the likely spot coverage of the star at the time of various observations by \textit{HST} and \textit{Spitzer}, and the resulting radius overestimation factor due to unocculted spots is calculated. Occulted spots are also accounted for. Whilst this method still relies on estimating the (non-zero) spot coverage when the stellar flux is at its peak, it represents a vast improvement on methods that make no attempt to account for star spots. It also helps when stitching together measurements from multiple instruments as it implicitly corrects for baseline offsets due to stellar activity. 

Of course, it is not possible to monitor the activity of every potential target star for \textit{JWST}, but for the most interesting targets around active stars it is an approach that must be considered. It relies on the availability of small, ground-based telescopes such as APT. 

\subsection{Target searches}
In order to fully exploit \textit{JWST}'s sensitivity, we require more super-Earth targets orbiting relatively nearby M dwarf stars. Over the next few years, ground-based and space-based surveys should provide more of these interesting targets. The \textit{Next Generation Transit Survey (NGTS)} \citep{chazelas12}
 is particularly suited to this because it operates between 600 and 900 nm, a wavelength range closer to the flux peak of cooler stars. Ideally, \textit{JWST} would undertake an atmospheric survey of a range of super-Earth planets similar to GJ 1214b, to investigate the typical range of atmospheric types and conditions, but at the moment there are very few for which follow up is possible. \textit{JWST}'s high sensitivity may make follow up easier for more distant targets, but if the majority of these small, super-Earth worlds are as cloudy as GJ 1214b, atmospheric signals for planets around more distant stars are still likely to be prohibitively small. Therefore, an important goal for the few years leading up to the launch of \textit{JWST} is the detection of other super-Earth planets at comparable distances to that of GJ 1214b. The MEarth survey that discovered GJ 1214b and its counterpart, MEarth South \citep{jirwin14}, are likely to contribute to \textit{JWST}'s target list of small planets, as will the K2 mission \citep{demory13,vanderburg14} and SPECULOOS survey \citep{gillon13}. The addition of the HAT-South network \citep{bakos09} to the radial velocity search effort should also yield more planets with constrained masses.

Other surveys such as the \textit{Transiting Exoplanet Survey Satellite}  (\textit{TESS}, \citealt{ricker14}), due to launch in 2017 to look at nearby bright stars, and continuing efforts such as the WASP survey for hot Jupiters \citep{christian06} will also provide interesting targets for \textit{JWST}.

\subsection{Cloudy atmosphere models}
Clouds clearly play a very important role in the atmosphere of GJ 1214b, and there is also evidence of their presence in hot Jupiter atmospheres. Until recently, the majority of atmospheric models have been cloud-free, but as the evidence for exoplanet clouds increases this is no longer a reasonable approach. \textit{JWST} transit observations should aid discrimination between cloudy and clear atmospheres, and for favourable targets may allow constraints to be placed on the composition, structure and particle size of any clouds present (e.g. \citealt{wakeford14}). We defer more detailed explanation of this issue to a future paper, which will focus explicitly on determing cloud properties using \textit{JWST} transmission spectra.

\subsection{Other instruments and observing scenarios}
In this paper, we have considered the instruments and modes which allow us to maximise the wavelength coverage obtainable with \textit{JWST} for the smallest number of separate transit observations. Other observational scenarios are possible, and may indeed be preferable for secondary transit or for less active stars. Two other near-infrared instruments, NIRISS and NIRCam, will also have the capability to do transit spectroscopy, although neither can capture the 0.6---5 $\upmu$m wavelength region simultaneously. The NIRISS slitless spectroscopy mode covers 0.7---2.5 $\upmu$m and is specifically designed for transit spectroscopy, so this will be a useful tool for capturing that specific wavelength range. NIRCam offers the opportunity for slitless spectroscopy between 2.5 and 5 $\upmu$m, at R~2000. 

NIRSpec and MIRI can both be used at higher resolving power than considered here, and whilst this also limits the instantaneous spectral range achievable, these modes will no doubt be useful. For example, higher resolving power can resolve the shapes of bands in secondary transit sufficiently to provide constraints on vertical variation of trace species, or probe narrow line cores above the cloud level in primary transit that would otherwise be missed. The MIRI integral field unit's greater wavelength range provides the potential to place stronger constraints on gases such as CO$_2$ that have significant bands longwards of 12 $\upmu$m, and may also be useful for secondary transits of cooler planets where their flux may be expected to peak. The bright target limit for NIRSpec is also less stringent when using higher resolution modes, as the photons are spread out more across the detector, avoiding problems of saturation. In future work we aim to consider some of these observations; however, we stress that stitching together more segments to make a coherent spectrum will require a better understanding of systematics. 

\section{Conclusion}
\textit{JWST} will be a powerful tool for exoplanet transit spectroscopy of a wide range of planets, due to its broad wavelength coverage and high sensitivity. This is the first space telescope built in the exoplanet spectroscopy era, and so for the first time we have the opportunity to test and characterise detectors prior to launch with a view to using them for transit spectroscopy. This should allow us to achieve the required precision of ~10---100 ppm in transit depth.

\textit{JWST} will allow us to characterise the temperature structure and atmospheric composition of hot Jupiters orbiting bright stars, with gas abundances retrieved down to 0.5 ppmv even for systems at large distances, ensuring a large range of accessible targets. It will also improve our understanding of cloudy super-Earths such as GJ 1214b by providing more wavelength coverage to allow distinction between different cloud models, and would enable us to determine some properties of Earth-like planets around cool stars provided they are close enough. However, accurate retrievals for many of these objects will depend on either observing relatively inactive stars, or using stellar monitoring as in \citet{pont13} to correct for the presence of starspots. For high SNR targets such as the hot Neptune-M dwarf system considered here, stellar activity is likely to be the limiting factor on how accurately we can constrain the atmosphere, as a 10\% star spot coverage can lead to errors of up to an order of magnitude in the H$_2$O abundance retrieval. 

\section{Acknowledgements}
JKB and PGJI acknowledge the support of the Science and Technology Facilities Council for this research. LNF is supported by a Royal Society University Research Fellowship. We thank the anonymous reviewer for some very helpful and constructive comments. 

\bibliographystyle{mn2e}
\bibliography{bibliography}

\label{lastpage}
\end{document}